\documentclass[]{pasj01}
\draft

\begin{document} 
\Received{}
\Accepted{}

\title{Asymmetric Expansion of the Fe ejecta in Kepler's Supernova Remnant}

\author{Tomoaki \textsc{kasuga}\altaffilmark{1}}
\altaffiltext{1}{Department of Physics, Graduate school of Science, The University of Tokyo, 7-3-1 Hongo, Bunkyo-ku, Tokyo 113-0033, Japan}
\email{kasuga@juno.phys.s.u-tokyo.ac.jp}

\author{Toshiki \textsc{sato}\altaffilmark{2}}
\altaffiltext{2}{RIKEN, 2-1 Hirosawa, Wako, Saitama 351-0198, Japan}

\author{Koji \textsc{mori}\altaffilmark{3}}
\altaffiltext{3}{Department of Applied Physics, University of Miyazaki, 1-1 Gakuen Kibana-dai Nishi, Miyazaki 889-2192, Japan}

\author{Hiroya \textsc{yamaguchi}\altaffilmark{4, 5}}
\altaffiltext{4}{NASA Goddard Space Flight Center, Code 662, Greenbelt, MD 20771, USA}
\altaffiltext{5}{Department of Astronomy, University of Maryland, College Park, MD 20742, USA}

\author{Aya \textsc{bamba}\altaffilmark{1, 6}}
\altaffiltext{6}{Research Center for the Early Universe, School of Science, The University of Tokyo, 7-3-1 Hongo, Bunkyo-ku, Tokyo 113-0033, Japan}


\maketitle

\begin{abstract}
The ejecta kinematics of supernova remnants (SNRs) is one of crucial clues to understand the explosion mechanism of type Ia supernovae (SNe). 
In particular, the kinematic asymmetry of iron-peak elements provides the key to understanding physical processes taking place in the core of the exploding white dwarfs (WDs) 
although it has been poorly understood by observations. 
In this paper, we show for the first time the asymmetric expansion structure in the line-of-sight direction of Fe ejecta in Kepler's SNR revealed by spectral and imaging analysis using the Chandra archival data. 
We found that the K$\alpha$ line centroid energy and line width is relatively lower ($<$ 6.4 keV) and narrower ($\sim$80 eV) around the center of the remnant, 
which implies that the majority of the Fe ejecta in the central region is red-shifted. 
At the outer regions, we identify bright blue-shifted structures as it might have been ejected as high velocity dense clumps. 
Taking into account the broad population of the Fe charge states, 
we estimate the red-shifted velocity of $\sim$2,000 km s$^{-1}$ and the blue-shifted velocity of $\sim$3,000 km s$^{-1}$ for each velocity structure. 
We also present a possibility that a portion of the Fe ejecta near the center are interacting with the dense circumstellar medium (CSM) on the near side of the remnant. 
As the origin of the asymmetric motion of the Fe ejecta, we suggest three scenarios; 
(1) the asymmetric distribution of the CSM, 
(2) the ``shadow'' in Fe cast by the companion star, 
and (3) the asymmetric explosion.
\end{abstract}

\section{Introduction}\label{sec:intro}

Type Ia supernovae (SNe), widely believed to result from thermonuclear explosions of white dwarfs, are particularly 
important phenomena in the universe, because of their role as distance indicators (i.e., standardizable candle) in cosmology 
and major sources of the Fe group elements. However, many of their fundamental aspects, such as how their progenitors 
evolve and explode, are still poorly understood (e.g., \cite{Maoz14,Maeda16}).
X-ray observations of supernova remnants (SNRs) offer a unique way to address the open questions regarding 
the progenitor's characteristics through investigations of their elemental composition and dynamics.

Kepler's SNR (SN 1604) is the youngest historically-recorded type Ia SNR in our galaxy (\cite{Vink17}). 
Its progenitor nature has been extensively studied from many aspects, including the post-explosion light curve 
(\cite{Baade43}) as well as spectral and morphological properties of the remnant (e.g., \cite{Reynolds07,Park13}). 
Previous X-ray studies revealed the presence of dense, asymmetric circumstellar medium (CSM) at a distance of a few parsecs 
from the explosion site (e.g., \cite{Blair07,Reynolds07,Williams12,Burkey13,Katsuda15}). 
This implies a so-called ``single-degenerate'' progenitor system (\cite{Whelan73}) as the origin of Kepler's SNR, 
although no surviving companion or light echo has been detected (\cite{Kerzendorf14,Sato17b}). 
The characteristics of progenitor's explosion and environment have been studied from the SNR dynamics as well. 
Thanks to the excellent angular resolution of Hubble Space Telescope and Chandra, the proper motion of 
the blast wave on the projected plane was accurately measured (\cite{Sankrit05,Katsuda08,Vink08,Sankrit16}), 
whose typical velocity is $\sim$\timeform{0.08''} yr$^{-1}$ (\cite{Sankrit16}). 
However, the distance to this SNR is poorly constrained (e.g., 4.4--5.9 kpc: \cite{Sankrit16}), 
causing a substantial uncertainty in the expansion velocity. 

A spectroscopic study of the line-of-sight Doppler velocity is, on the other hand, not subject to such uncertainties and thus offers direct clues to the SNR dynamics. 
Suzaku observations of Tycho's SNR (another young type Ia supernova remnant) revealed that the widths of Si and Fe emission lines increase toward the SNR center, 
indicating evidence for ejecta expansion along the line of sight (\cite{Furuzawa09,Hayato10}).
More recently, \citet{Williams17} and \citet{Sato17a} independently analyzed high-resolution Chandra data and found 
that each of small ejecta clumps have different line-of-sight velocities, 
i.e., their emission lines are either red- or blue-shifted.
For Kepler's SNR, \citet{Sato17b} determined three-dimensional velocity of intermediate-mass elements (e.g., Si, S, Ar) 
by combining the proper-motion and Doppler-shift measurements using Chandra data of multiple epochs. 
Using the direct velocity measurements, 
they found a wide range of the ejecta velocity, from $\sim$1,000 km s$^{-1}$ (substantially decelerated) to $\sim$10,000 km s$^{-1}$ (almost freely expanding). 
The study newly suggested the past activities of the progenitor system that could structure the ambient medium, 
including regions both of higher and of lower density. 
On the other hand, three-dimensional dynamics of other iron-group elements have not been understood clearly yet.
Since Fe originate from hotter regions of an exploding white dwarf than where the Si ejecta were generated (e.g., \cite{Iwamoto99}), 
their distributions in both spatial and velocity spaces strongly constrain the SN explosion characteristics.

Recently, multi-dimensional explosion simulations have suggested that type Ia SNe may be highly asymmetric (\cite{Livne05,Kuhlen06,Kasen09}). 
Also, based on observations of velocity shifts in late-phase nebular spectra, 
it was argued that type Ia SNe may result from asymmetric explosions (e.g., \cite{Maeda10}). 
Such an asymmetric explosion is thought to cause an asymmetric distribution of $^{56}$Ni (which decays to $^{56}$Fe), 
so it might be seen in the Fe distribution in SNRs.
In case of Kepler's SNR, asymmetrically-distributed Fe emissions toward its interior have been already indicated (\cite{Cassam04,Burkey13}).
It is notable that the asymmetric distribution of the Fe emissions is thought to be not due to the asymmetric distribution of the CSM 
because the Fe K distribution does not coincide with the CSM distribution (\cite{Burkey13}).
Therefore, the kinetic asymmetry in Fe ejecta may be directly related to the explosion mechanism. 
In order to examine such asymmetry in Kepler's SNR, 
we focus on the Fe ejecta and perform the first measurement of their line-of-sight velocity structure in the small spatial scale 
by utilizing the superb angular resolution of Chandra. 

In section \ref{sec:obs}, we describe observations and data reduction. 
The results of data analysis are given in section \ref{sec:analysis}.
We discuss the obtained results in section \ref{sec:discussion} and conclude this work in section \ref{sec:conclusion}.

\section{Data selection and reduction}\label{sec:obs}
\begin{table}[!h]
  \tbl{Observation Log of Kepler's SNR.}{
  \begin{tabular}{cclr}
      \hline
      Obs. ID & (R.A., Decl.) & Obs. Start & Exposure \\
       & & & (ks)\\ 
      \hline
      6714 & (\timeform{17h30m42.00s}, \timeform{-21D29'00.00''}) & 2006 Apr 27 & 157.8 \\
      6716 & (\timeform{17h30m42.00s}, \timeform{-21D29'00.00''}) & 2006 May 05 & 158.0\\
      6717 & (\timeform{17h30m41.24s}, \timeform{-21D29'31.45''}) & 2006 Jul 17 & 106.8\\
      7366 & (\timeform{17h30m41.24s}, \timeform{-21D29'31.45''}) & 2006 Jul 17 & 51.5\\
      6718 & (\timeform{17h30m41.24s}, \timeform{-21D29'31.45''}) & 2006 Jul 24 & 107.8\\
      6715 & (\timeform{17h30m41.24s}, \timeform{-21D29'31.45''}) & 2006 Aug 03 & 159.1\\
      \hline
    \end{tabular}}\label{tab:chandralog}
\begin{tabnote}
\end{tabnote}
\end{table}

\noindent The Chandra X-ray Observatory has observed Kepler's SNR several times using the Advanced CCD Imaging 
Spectrometer Spectroscopic-array (\cite{Bautz98}). 
Since our aim is to determine the accurate line-of-sight velocity in small regions, 
we use only the deepest observation dataset obtained in 2006 with the total effective exposure of 741.0 ks as listed in table \ref{tab:chandralog}
to avoid the effect of the proper motion. 
We reprocessed the event data in each obs ID using the {\tt chandra\_repro} task in the {\tt CIAO 4.9} software package with {\tt CALDB 4.7.4}. 
For spectral analysis in section \ref{sec:radial} and \ref{sec:map}, 
we made spectra, response and ARF files using {\tt specextract} for each observation, 
and combined them using the {\tt combine\_spectra} script.
For image analysis in section \ref{sec:map}, 
we merged raw data at first using {\tt merge\_obs}  to improve the photon statistics. 
\citet{Sato17b} reported that the shift of astrometric alignment is less than \timeform{0.35''}, 
much smaller than the region size we are interested in. 
We thus made no correction on the absolute astrometry in our analysis.

\section{Analysis and result}\label{sec:analysis}
\subsection{Radial profile}\label{sec:radial}
\begin{figure}[!h]
 \begin{center}
  \includegraphics[width=8cm]{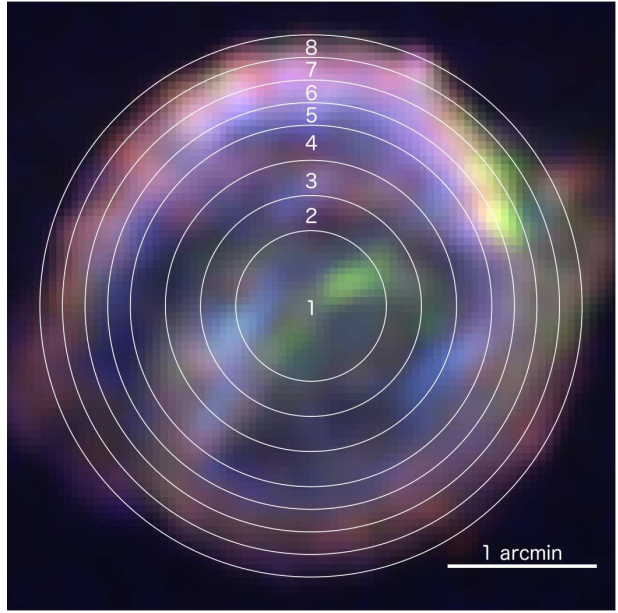}
 \end{center}
\caption{Chandra three-color image of Kepler's SNR, binned with 3 arcsec and smoothed with a Gaussian kernel of 9 arcsec. The scale is in square root. North is up and east is to the left. The red, green and blue images are made using the energy bands containing Si K emission (1.78--1.93 keV), O K emission (0.50--0.70 keV) and Fe K emission (6.20--6.70 keV), respectively. The white circles indicate where we extract spectra to generate the radial profiles in figure \ref{fig:result_radial}.} \label{fig:region_radial}
\end{figure}

\begin{figure}[!h]
 \begin{center}
  \includegraphics[width=8cm]{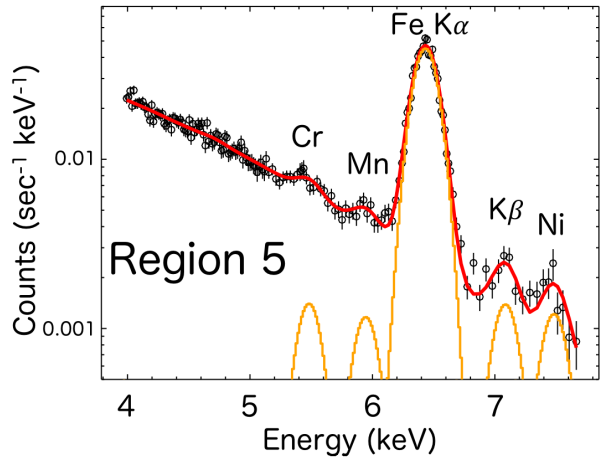}
 \end{center}
\caption{Example spectrum in the 4.0--7.7 keV band, extracted from the brightest annulus Region 5. 
The red line represents the best-fit model, and orange Gaussians indicate the contribution of Cr, Mn, Fe, and Ni K$\alpha$ emission, and Fe K$\beta$ emission.}\label{fig:spec}
\end{figure}

\begin{figure*}[!t]
 \begin{center}
  \includegraphics[width=16cm]{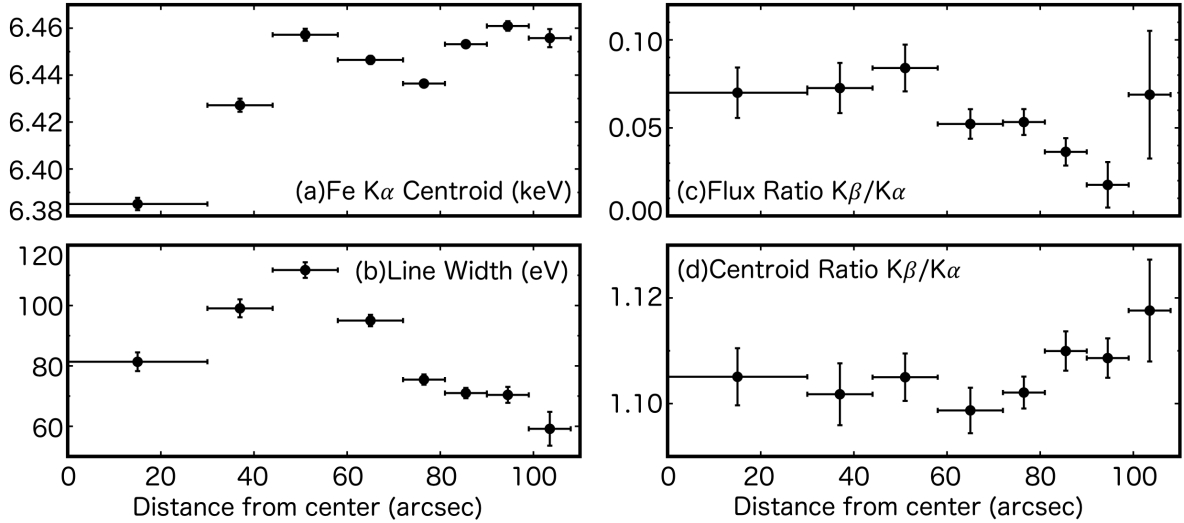}
 \end{center}
\caption{The radial profiles of (a) Fe K$\alpha$ centroid, (b) 1$\sigma$ line width, (c) K$\beta$/K$\alpha$ flux ratio and (d) K$\beta$/K$\alpha$ centroid energy ratio. Error bars represent the 1$\sigma$ confidence level.}\label{fig:result_radial}
\end{figure*}

Figure \ref{fig:region_radial} shows a three-color image of Kepler's SNR, 
where red, green and blue indicate K-shell emission band of Si, O and Fe, respectively. 
Thanks to the fine point spread function (half power diameter $\sim$\timeform{0.5''}) of the High Resolution Mirror Assembly, 
Chandra successfully resolved a number of small-scale ejecta clumps with different chemical compositions (e.g., \cite{Reynolds07}). 
Nevertheless, we first investigate one-dimensional radial profiles of the surface brightness, centroid energy, 
and line width of the Fe K$\alpha$ emission in order to investigate the overall trend. 
We divide the entire SNR into 8 annulus regions (Regions 1--8), 
assuming the SNR center of (\timeform{17h30m41.321s}, \timeform{-21D29'30.510''}) (\cite{Sato17b}). 
The radius of Region 1 is \timeform{30''} 
and the widths of the outer annulus are \timeform{14''} (Regions 2--4) or \timeform{9''} (Regions 5--8).
The 4.0--7.7 keV spectrum of the brightest annulus (Region 5) is shown in figure \ref{fig:spec}. 
This energy band contains K$\alpha$ emission of Cr, Mn, Fe, and Ni, and Fe K$\beta$ emission (\cite{Park13}), 
which are also confirmed in our Chandra spectrum. 
Therefore, we fit the spectrum with Gaussian models for the five line features plus a power-law continuum.
The spectra from all eight regions are simultaneously fitted to link the centroid of Cr, Mn and Ni 
among the regions and the Fe K$\beta$ centroid between Regions 6 and 7. 
All the other parameters (centroid, width, and brightness) for these lines were treated as free parameters among the regions. 
The fitting was successful with parameters listed in table \ref{tab:result_radial} with $\chi^2$ (d.o.f.) of 0.93371 (1089), 
and the resulting radial profiles are given in figure \ref{fig:result_radial}.

\begin{table*}[!ht]
  \tbl{Best-fit spectral parameters for the radial profile.}{
  \begin{tabular}{cccccccc}
      \hline
      Line & Centroid & Width ($1\sigma$) & Brightness\footnotemark[$*$] & & Centroid & Width ($1\sigma$)& Brightness\footnotemark[$*$]\\
       & (keV) & (eV) & & & (keV) & (eV) & \\
      \hline
       & \multicolumn{3}{c}{region 1} & & \multicolumn{3}{c}{region 2} \\
      \cline{2-4} \cline{6-8}
      Cr K$\alpha$	& $5.484 \pm 0.014$ & = Fe K$\alpha$ & $1.0 \pm 0.7$ &		& $\dagger$ & = Fe K$\alpha$ & $1.7 \pm 0.7$ \\
      Mn K$\alpha$	& $5.948 \pm 0.015$ & = Fe K$\alpha$ & $1.3 \pm 0.8$ &		& $\ddagger$ & = Fe K$\alpha$ & $1.8 \pm 0.8$ \\
      Fe K$\alpha$	& $6.385 \pm 0.003$ & $81 \pm 3$ & $94.7 \pm 2.2$ &		& $6.427 \pm 0.003$ & $99 \pm 3$ & $95.6 \pm 2.1$ \\
      Fe K$\beta$	& $7.056 \pm 0.034$ & = Fe K$\alpha$ & $6.6 \pm 1.4$ &		& $7.081 \pm 0.038$ & = Fe K$\alpha$ & $7.0 \pm 1.4$ \\
      Mn K$\alpha$	& $7.499 \pm 0.012$ & = Fe K$\alpha$ & $4.1 \pm 2.1$ &		& $\S$ & = Fe K$\alpha$ & $1.05 \pm 2.2$ \\
       & \multicolumn{3}{c}{region 3} & & \multicolumn{3}{c}{region 4} \\
      \cline{2-4} \cline{6-8}
      Cr K$\alpha$	& $\dagger$ & = Fe K$\alpha$ & $1.7 \pm 0.7$ &			& $\dagger$ & = Fe K$\alpha$ & $1.5 \pm 0.6$ \\
      Mn K$\alpha$	& $\ddagger$ & = Fe K$\alpha$ & $2.9 \pm 0.7$ &			& $\ddagger$ & = Fe K$\alpha$ & $1.3 \pm 0.7$ \\
      Fe K$\alpha$	& $6.457 \pm 0.003$ & $112 \pm 3$ & $99.7 \pm 1.9$ &		& $6.446 \pm 0.002$ & $95 \pm 2$ & $133 \pm 2$ \\
      Fe K$\beta$	& $7.135 \pm 0.029$ & = Fe K$\alpha$ & $8.4 \pm 1.3$ &		& $7.083 \pm 0.028$ & = Fe K$\alpha$ & $7.0 \pm 1.1$ \\
      Mn K$\alpha$	& $\S$ & = Fe K$\alpha$ & $7.8 \pm 2.0$ &				& $\S$ & = Fe K$\alpha$ & $10.4 \pm 1.7$ \\
       & \multicolumn{3}{c}{region 5} & & \multicolumn{3}{c}{region 6} \\
      \cline{2-4} \cline{6-8}
      Cr K$\alpha$	& $\dagger$ & = Fe K$\alpha$ & $3.2 \pm 0.8$ &			& $\dagger$ & = Fe K$\alpha$ & $3.0 \pm 0.7$ \\
      Mn K$\alpha$	& $\ddagger$ & = Fe K$\alpha$ & $3.5 \pm 0.8$ &			& $\ddagger$ & = Fe K$\alpha$ & $4.1 \pm 0.8$ \\
      Fe K$\alpha$	& $6.436 \pm 0.001$ & $75 \pm 2$ & $181 \pm 2$ &			& $6.453 \pm 0.001$ & $71 \pm 2$ & $159 \pm 2$ \\
      Fe K$\beta$	& $7.094 \pm 0.019$ & = Fe K$\alpha$ & $9.6 \pm 1.3$ &		& $7.163 \pm 0.024$ & = Fe K$\alpha$ & $5.8 \pm 1.2$ \\
      Mn K$\alpha$	& $\S$ & = Fe K$\alpha$ & $12.2 \pm 2.0$ &				& $\S$ & = Fe K$\alpha$ & $13.3 \pm 1.9$ \\
      & \multicolumn{3}{c}{region 7} & & \multicolumn{3}{c}{region 8} \\
      \cline{2-4} \cline{6-8}
      Cr K$\alpha$	& $\dagger$ & = Fe K$\alpha$ & $3.0 \pm 0.7$ &			& $\dagger$ & = Fe K$\alpha$ & $0.2$~\footnotemark[$\#$] \\
      Mn K$\alpha$	& $\ddagger$ & = Fe K$\alpha$ & $2.0 \pm 0.7$ &			& $\ddagger$ & = Fe K$\alpha$ & $0.8 \pm 0.5$ \\
      Fe K$\alpha$	& $6.461 \pm 0.002$ & $70 \pm 3$ & $83.8 \pm 1.7$ &		& $6.456 \pm 0.004$ & $71 \pm 2$ & $26.1 \pm 1.0$ \\
      Fe K$\beta$	& $\|$ & = Fe K$\alpha$ & $1.5 \pm 1.1$ &				& $7.215 \pm 0.062$ & = Fe K$\alpha$ & $1.8 \pm 0.9$ \\
      Mn K$\alpha$	& $\S$ & = Fe K$\alpha$ & $6.4 \pm 1.6$ &				& $\S$ & = Fe K$\alpha$ & $4.2 \pm 1.4$ \\
     \hline
    \end{tabular}}\label{tab:result_radial}
\begin{tabnote}
Errors are at a 1$\sigma$ confidence level.\\
\footnotemark[$*$] Unit is $\times 10^{-10}$ cm$^{-2}$ s$^{-1}$ arcsec$^{-2}$.\\
\footnotemark[$\dagger$] \footnotemark[$\ddagger$] \footnotemark[$\S$] Cr, Mn, and Ni K$\alpha$ centroid is linked with that in region 1. \\
\footnotemark[$\|$] Fe K$\beta$ centroid in region 7 is linked with that in region 6. \\
\footnotemark[$\#$] Error range of the Cr K$\alpha$ brightness in region 8 is not constrained.
\end{tabnote}
\end{table*}

The lowest Fe K$\alpha$ centroid energy is found at the innermost region (figure \ref{fig:result_radial}(a)), which is somewhat unexpected. 
Moreover, the measured value (6.385 $\pm$ 0.003 keV) is even lower than the theoretical energy of the K$\alpha$ fluorescence from neutral Fe. 
This suggests that the Fe emission from the SNR center is 
substantially red-shifted due to the asymmetric distribution of the line-of-sight velocity. 
This interpretation is also supported by the radial profile of the line width (figure \ref{fig:result_radial}(b)). 
A relatively narrow emission found at the innermost region is in contrast to Tycho's SNR, where the largest line width 
is confirmed at the SNR center as expected for a uniformly expanding shell  (\cite{Furuzawa09,Hayato10,Sato17a}). 
We also reveal radial trends in the Fe K$\beta$/K$\alpha$ flux and centroid ratios, which respectively show lower and higher 
values at the outer regions (figure \ref{fig:result_radial}(c) and (d)). 
Details will be discussed in section \ref{sec:ionization}.

In the case of Kepler's SNR, 
the reverse shock dynamics on the north and south side can be different from each other due to the biased-CSM distribution (e.g., \cite{Williams12}). 
It would make a difference of radial dependence of the ionization states (K$\beta$/K$\alpha$ ratios) between the north and south sides. 
However, we did not find such a significant difference in radial profiles between north and south (see the Appendix). 
We therefore assume the same ionization states in one radial profile hereafter. 

\subsection{Small-scale velocity structure}\label{sec:map}

\begin{figure*}[!t]
 \begin{center}
  \includegraphics[width=16cm]{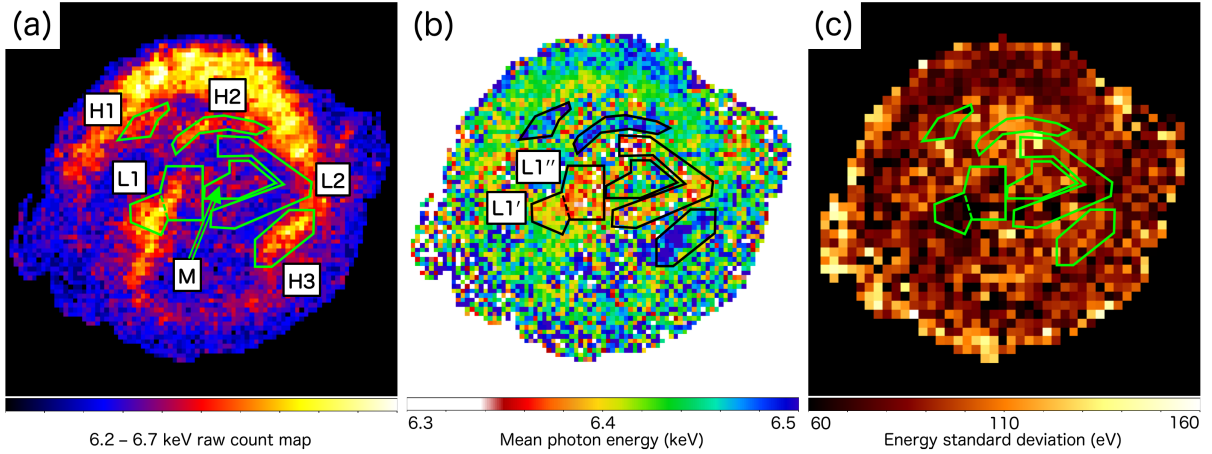}
 \end{center}
\caption{(a) Photon count, (b) mean photon energy ($\bar{E}$), and (c) standard deviation ($S$) maps in the 6.2--6.7 keV band. Each image bin corresponds to 3$\times$3 arcsec$^2$ for panels (a) and (b), and 6$\times$6 arcsec$^2$ for panel (c). Region names for section \ref{sec:map} are also shown.}\label{fig:F_M_S}
\end{figure*}

To investigate smaller-scale variations in the line centroid energy, we generate a mean photon energy (MPE) map, 
as \citep{Sato17a} applied to the Si K emission from Tycho's SNR. 
We first generate narrow band images (photon count maps) in every 20 eV between 6.2 keV and 6.7 keV and 
then calculate the MPE $\bar{E}$ for each pixel as
\begin{equation}
\bar{E} \equiv \frac{\sum_i n_i E_i}{\sum_i n_i} \ ,
\label{eq:E}
\end{equation}
where $n_i$ and $E_i$ are the number of photons and average energy in each energy grid, respectively.
The resulting MPE map is shown in figure \ref{fig:F_M_S}(b),  
where an average photon count in a single pixel $\sim$25. 
In line with the 1D radial profile (section \ref{sec:radial}), the lowest MPE is found near the SNR center. 
However, the velocity structure is found to be more complex. 
The red-shifted (low MPE) pixels appear to be relatively diffuse and located near the center 
whereas the blue-shifted (high MPE) pixels appear to be patchy and distributed in the outer regions. 
In particular, the red-shifted pixels from an annular shape does not coincide with the Fe K$\alpha$ flux distribution (figure \ref{fig:F_M_S}(a)). 
We also discover several blue-shifted regions, that are prominent in the flux image as well. 

We also calculate standard deviation $S$ of the Fe K$\alpha$ photon energies in each pixel 
to investigate two-dimensional line width distribution:  
\begin{eqnarray}
S &\equiv& \sqrt{\frac{\sum_i n_i (E_i - \bar{E})^2}{\sum_i n_i}} = \sqrt{\frac{\sum_i n_i E_i^2}{\sum_i n_i} - (\bar{E})^2} \ . 
\label{eq:S}
\end{eqnarray}
The result is shown in figure \ref{fig:F_M_S}(c). 
This is also consistent with the radial profile of the line width (section \ref{sec:radial}). 
The smallest standard deviation is found at the outer layer (where the line-of-sight velocity is expected to be low) 
and the central regions (where the Fe K$\alpha$ emission is significantly red-shifted).
We also find that the deviation is generally small in the southeast quadrant, 
compared to the other regions in the SNR.
Note that the MPEs at the outermost regions should not be interpreted as the Fe K$\alpha$ centroid, 
because these regions are dominated by strong synchrotron X-rays (\cite{Bamba05}).

\begin{figure}[!h]
 \begin{center}
  \includegraphics[width=8cm]{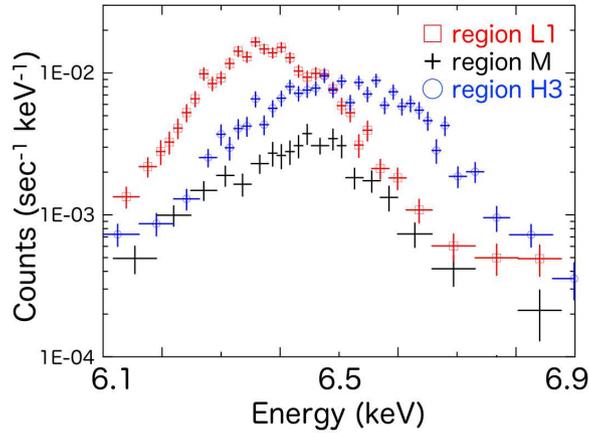}
 \end{center}
\caption{Magnified spectra around the Fe K$\alpha$ emission (6.1--6.9 keV) of Regions L1 (red), M (black), and H3 (blue).}\label{fig:region_detail_spec}
\end{figure}

For more quantitative study, we extract spectra from six characteristic regions indicated in figure \ref{fig:F_M_S}: 
high-energy regions (H1--H3), low-energy/narrow-line regions (L1, L2), and a region surrounded by the low-energy regions (M). 
The Fe K$\alpha$ spectra of Region H3, L1, and M are compared in figure \ref{fig:region_detail_spec}. 
We fit the 4.0--7.7 keV spectrum of each region with Gaussians (for the emission lines) and a power law (for the continuum), 
obtaining the best-fit values given in table \ref{tab:result_detail}. 
The results are generally consistent with the MPE and deviation maps; 
e.g., the Fe K$\alpha$ lines of Region L1 and L2 are indeed lower than 6.4 keV and narrow.

\begin{table*}[!h]
  \tbl{Best-fit spectral parameters for small characteristic regions.}{
  \begin{tabular}{cccccccccc}
      \hline
      Region & Centroid & Width ($1\sigma$) & Norm\footnotemark[$*$] & Brightness\footnotemark[$\dagger$] & & Centroid & Width ($1\sigma$)& Norm\footnotemark[$*$] & Brightness\footnotemark[$\dagger$]\\
       & (keV) & (eV) & & & & (keV) & (eV) & & \\
      \hline
       & \multicolumn{4}{c}{Fe K$\alpha$} & & \multicolumn{4}{c}{Fe K$\beta$} \\
      \cline{2-5} \cline{7-10}
      H1	& $6.483 \pm 0.007$ & $83 \pm 8$	 & $58 \pm 3$ & $166 \pm 9$ &		& $7.285 \pm 0.073$ & = Fe K$\alpha$ & $5 \pm 2$ & $14 \pm 6$ \\
      H2	& $6.496 \pm 0.007$ & $102 \pm 9$ & $86 \pm 4$ & $165 \pm 8$ &		& $7.185 \pm 0.087$ & = Fe K$\alpha$ & $4 \pm 3$ & $8 \pm 6$ \\
      H3	& $6.505 \pm 0.004$ & $108 \pm 4$ & $179 \pm 5$ & $200 \pm 6$ &	& $7.204$	\footnotemark[$\ddagger$]  & = Fe K$\alpha$ & $5 \pm 3$ & $6 \pm 3$ \\
      L1	& $6.381 \pm 0.003$ & $69 \pm 3$	 & $215 \pm 5$ & $161 \pm 4$ &	& $7.044 \pm 0.051$ & = Fe K$\alpha$ & $7 \pm 3$ & $5 \pm 2$ \\
      L2	& $6.388 \pm 0.005$ & $86 \pm 6$ & $105 \pm 4$ & $66 \pm 3$ &		& $7.001 \pm 0.093$ & = Fe K$\alpha$ & $7 \pm 3$ & $4 \pm 2$ \\
      M	& $6.438 \pm 0.008$ & $94 \pm 9$	 & $55 \pm 3$ & $80 \pm 4$ &		& $7.056 \pm 0.213$ & = Fe K$\alpha$ & $2 \pm 2$ & $3 \pm 3$ \\
     \hline
    \end{tabular}}\label{tab:result_detail}
\begin{tabnote}
Errors are at a 1$\sigma$ confidence level.\\
\footnotemark[$*$] Units is $\times 10^{-7}$ cm$^{-2}$ s$^{-1}$.\\
\footnotemark[$\dagger$] Units is $\times 10^{-10}$ cm$^{-2}$ s$^{-1}$ arcsec$^{-2}$.\\
\footnotemark[$\ddagger$] Error range is not constrained.
\end{tabnote}
\end{table*}

\section{Discussion}\label{sec:discussion}

We have investigated both radial trend and small-scale distribution of the Fe K emission characteristics 
and spatially resolved the red- and blue-shifted ejecta components, for the first time. In this section, we aim to 
constrain the actual line-of-sight velocity of the Fe ejecta from the observed centroid energy in each small region.
We should note, however, that the K-shell fluorescence energy depends also on the charge number of Fe ions 
\citep{Yamaguchi14}. Therefore, we first investigate the charge population in the shocked ejecta 
to estimate the average centroid energy of Fe K$\alpha$ lines at the rest frame. For this purpose, 
we perform plasma diagnostics in section \ref{sec:ionization}, and subsequently determine the line-of-sight velocity in section \ref{sec:velocity}. 

\subsection{Ionization state of Fe ejecta}\label{sec:ionization}

\citet{Yamaguchi14} presented that the Fe K$\beta$/K$\alpha$ flux ratio is sensitive to the Fe charge number (or ionization degree), 
because the fluorescence yields of these lines depend on the number of bound electrons in the $2p$ and $3p$ shells. 
More specifically, strong K$\beta$ emission is expected only from low-ionization plasma where $3p$ electrons are still bound. 
We find in figure \ref{fig:result_radial}(c) that the Fe K$\beta$/K$\alpha$ flux ratio gradually decreases towards the outer regions, 
indicating that the average charge number is higher at the outer region. This result is consistent with the fact that 
the SNR reverse shock propagates inward so that the ejecta in the outer layer get ionized earlier. 
We also find that the Fe K$\beta$/K$\alpha$ centroid energy ratio, which is independent of the kinematic Doppler effect increases with the distance from the SNR center.
This trend is also due to the ionization effect; the fluorescence energy of Fe K$\beta$ emission increases faster than 
that of Fe K$\alpha$ emission with respect to the charge number (see Table 2 of \citet{Yamaguchi14}).  

\begin{figure}[!h]
 \begin{center}
  \includegraphics[width=8cm]{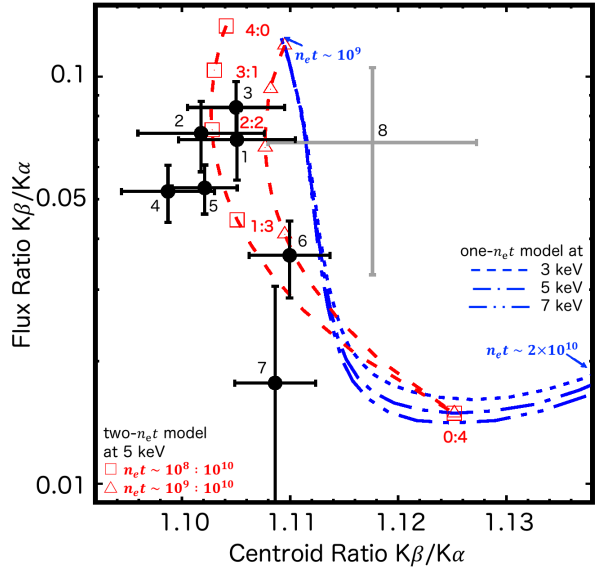}
 \end{center}
\caption{Relationship between the Fe K$\beta$/K$\alpha$ centroid ratio (horizontal axis) and flux ratio (vertical axis) measured in the annulus regions (Regions 1--8, indicated in the plot). The error bars represent 1$\sigma$ confidence. Blue curves indicate theoretically-expected relations for `single-$n_et$' plasmas with $n_et$ values ranging from $10^9$ cm$^{-3}$ s (upper-left side) to $2 \times 10^{10}$ cm$^{-3}$ s (bottom-right side). Various electron temperatures (3, 5, and 7 keV) are considered. Red dashed curves assume `two-$n_et$' models consisting of $n_{\mathrm{e}}t$ = $10^{8}$ and $10^{10}$ cm$^{-3}$ s plasmas (squares) or $n_{\mathrm{e}}t$ = $10^{9}$ and $10^{10}$ cm$^{-3}$ s plasmas (triangles). The fractions of each component (defined by the flux contributing to the Fe K$\alpha$ emission) are also indicated beside the symbols.}\label{fig:discussion_1}
\end{figure}

In figure \ref{fig:discussion_1}, the Fe K$\beta$/K$\alpha$ flux and centroid ratios observed in the eight annular regions (black crosses) 
are compared with the theoretical values calculated using the atomic data of \citet{Yamaguchi15}. The blue curves assume 
a single ionization age, $n_{\mathrm{e}}t$ (where $n_{\mathrm{e}}$ and $t$ are the electron density and the time elapsed 
since shock heating), ranging from $1 \times 10^9$ to $2 \times 10^{10}$ cm$^{-3}$\,s. 
The impact of different electron temperatures are explored for $kT_{\mathrm{e}} =$ 3--7 keV, 
which contains a previously-reported electron temperature of the Fe ejecta in this SNR ($\sim$5\,keV: \cite{Park13}). 
We find that the `one-$n_{\mathrm{e}}t$' models generally fail to reproduce the observed ratios. 
This implies the presence of a wider range of plasma conditions, because a lower K$\beta$/K$\alpha$ centroid ratio 
is expected if the K$\alpha$ and K$\beta$ emission is predominantly originate from the high- and low-ionized component, 
respectively, as in the case of Tycho's SNR \citep{Yamaguchi14}. 
We thus apply the `two-$n_{\mathrm{e}}t$' models consisting of $n_{\mathrm{e}}t$ = $10^{8}$ and $10^{10}$ cm$^{-3}$ s plasmas
or $n_{\mathrm{e}}t$ = $10^{9}$ and $10^{10}$ cm$^{-3}$ s plasmas with various fractions of each component. 
The red curves in figure \ref{fig:discussion_1} indicate the range of the calculated ratios, where a larger 
fraction of the low-ionized component is assumed at the higher upper-left (higher flux ratio) region. 
The observed Fe K properties are well explained with this `two-$n_{\mathrm{e}}t$' assumption. 

We find that the values observed in Regions 1--5 can be reproduced 
when the low-ionization component ($n_{\mathrm{e}}t$ of either $10^{8}$ or $10^{9}$ cm$^{-3}$ s) is responsible for 25\%--75\% of the Fe K$\alpha$ flux. 
On the other hand, the plasmas in Regions 6 -- 8 are dominated by the high-ionization component. 
It should be noted that the theoretical Fe K$\alpha$ centroid energy remains almost constant ($E \sim 6400$ eV) within the range of $n_{\mathrm{e}}t$ = $10^8$ and $10^9$ cm$^{-3}$ s 
and the difference in the K$\beta$/K$\alpha$ centroid ratios is mainly owing to the charge-number dependence of the K$\beta$ centroid energy \citep{Yamaguchi14}. 
Since the Fe K$\alpha$ centroid for the $n_{\mathrm{e}}t$ = $10^{10}$ cm$^{-3}$ s plasma is predicted to be $\sim 6450$ eV, 
we can estimate the rest frame centroid energy of the Fe K$\alpha$ emission from Regions 1--5 to be $(6400+6450)/2 \approx 6425$ eV, 
assuming the responsibility of the low-ionization component to be 50 \%. 
The estimated variation in the fraction of low-ionization component gives the uncertainty of $\sim$10 eV.

\subsection{Asymmetric structure of Fe ejecta motion}\label{sec:velocity}
All the small characteristic regions in figure \ref{fig:F_M_S}(a) are located in the annulus Region 1--4, 
where the Fe K$\alpha$ centroid in the ejecta-rest frame, $E_{0}$, is calculated to be 6.425 keV.
The centroid $E$ we observe is subject to the Doppler effect due to the line-of-sight velocity $v_\mathrm{sight}$;  
\begin{equation}
\frac{E_0-E}{E_0} = \frac{v_\mathrm{sight}}{c} \ ,
\end{equation}
where the positive $v_\mathrm{sight}$ represents red-shifted velocity and $c$ is the light velocity. 
The $v_\mathrm{sight}$ values derived at small regions is shown in table \ref{tab:velocity}. 
We also try dividing Region L1 into the eastern half (L1$^\prime$) and western half (L1$^{\prime \prime}$) and estimating the velocity (figure \ref{fig:F_M_S}(b)) 
under the assumption of same ionization state as that of region L1. 
We find no significant difference of velocity between two regions (also shown in table \ref{tab:velocity}). 
The statistics is not enough to discuss smaller regions than regions in figure \ref{fig:F_M_S}(a), 
so we still use Region L1 (not using Region L1$^\prime$ and L1$^{\prime \prime}$) in the following discussion. 

\begin{table}[!ht]
  \tbl{Ejecta velocity in small characteristic regions.}{
  \begin{tabular}{crc}
      \hline
      Region & $v_\mathrm{sight}$ \footnotemark[$*$]& Doppler shift \\
       & (km s$^{-1}$) & \\ 
      \hline
      H1 & $-2,700 \pm 700$ & blue-shift\\
      H2 & $-3,300 \pm 700$ & blue-shift\\
      H3 & $-3,700 \pm 700$ & blue-shift\\
      L1 & $2,100 \pm 700$ & red-shift\\
      ( L1' & $1,400 \pm 700$ & red-shift )\\
      ( L1'' & $2,700 \pm 700$ & red-shift )\\
      L2 & $1,700 \pm 700$ & red-shift\\
      M & $-600 \pm 800$ & no-shift\\
      \hline
    \end{tabular}}\label{tab:velocity}
\begin{tabnote}
 \footnotemark[$*$] The statistical and systematic errors are included. Positive velocity represents red-shifted ejecta.
\end{tabnote}
\end{table}

Region H1--H3 show significant blue-shift with the velocity of $\sim$3,000 km s$^{-1}$, 
whereas Region L1 and L2 are red-shifted with the velocity of $\sim$2,000 km s$^{-1}$. 
The apparent no-shift in Region M will be discussed later. 
The absolute values of the line-of-sight velocities of the blue-shifted ejecta in Region H1--H3 are relatively larger than those of the red-shifted ejecta in Region L1 and L2. 
In addition, Region H1--H3 are located at larger off-axis angles from the expansion center than Region L1 and L2, 
so that the difference between blue and red-shifted ejecta in three-dimensional (3D) space velocity may be even larger. 
The broader line width in these regions may also support the higher 3D space velocity of the blue-shifted ejecta. 
It is interesting to note that 
the blue-shifted ejecta in Region H1--H3 have a clear counterpart in the Fe K$\alpha$ line flux image (figure \ref{fig:F_M_S}(a)) 
whereas the red-shifted ejecta do not; 
the red portion in Region L1 in the MPE map (figure \ref{fig:F_M_S}(b)) does not appear to correspond to a bright structure running across the boundary of Region L1 toward the southeast in the flux image, 
and there is no characteristic structure in Region L2 in the flux image. 
The blue-shifted ejecta might have been ejected as high velocity dense clumps like the ``Fe knot'' in Tycho's SNR (\cite{Yamaguchi17}), 
while red-shifted ejecta might be an ensemble of relatively uniformly distributed smaller structures resulting in the diffuse appearance without distinct shape. 

\begin{figure}[!h]
 \begin{center}
  \includegraphics[width=8cm]{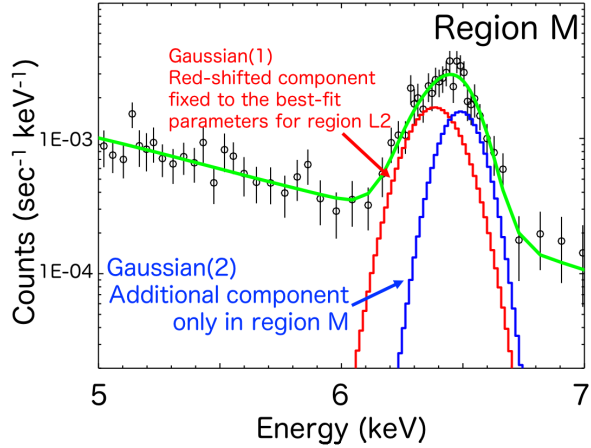}
 \end{center}
\caption{The 5--7 keV spectrum of Region M modeled with a power law and two Gaussians representing the red- and blue-shifted Fe K$\alpha$ emission. The line centroid and width of the former component are fixed to the best-fit values for Region L2.}\label{fig:discussion_2_E}
\end{figure}

\begin{table}[!ht]
  \tbl{Best-fit spectral parameters for the simultaneous fitting of Region L2 and M.}{
  \begin{tabular}{cccc}
      \hline
      Region\footnotemark[$*$] & Centroid & Width(1$\sigma$) & Norm\footnotemark[$\dagger$] \\
       & (keV) & (eV) & \\ 
      \hline
      L2	& $6.388 \pm 0.005$ & $86 \pm 6$ & $105 \pm 4$ \\
      M(1)	& =region L2 & =region L2 & $30 \pm 7$ \\
      M(2)	& $6.492 \pm 0.019$ & $53 \pm 25$ & $24 \pm 6$ \\
      \hline
    \end{tabular}}\label{tab:discussion_2_E}
\begin{tabnote}
 Errors indicate the 1$\sigma$ confidence limits. \\
 \footnotemark[$*$] M(1) and M(2) represent Gaussian 1 and 2, respectively. \\
 \footnotemark[$\dagger$] Units in $\times 10^{-7}$ cm$^{-2}$ sec$^{-1}$.
\end{tabnote}
\end{table}

\begin{figure}[!h]
 \begin{center}
  \includegraphics[width=8cm]{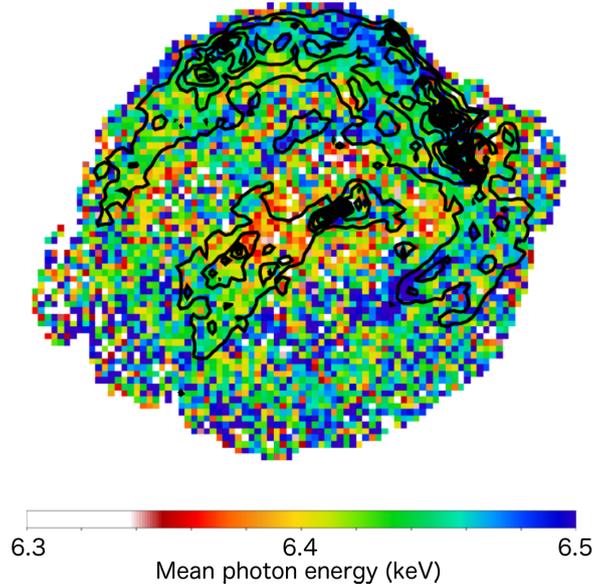}
 \end{center}
\caption{The same color map as figure \ref{fig:F_M_S}(b), where the Oxygen band flux image is overplotted in black contours. }\label{fig:discussion_2_map}
\end{figure}

\begin{figure}[!ht]
 \begin{center}
  \includegraphics[width=8cm]{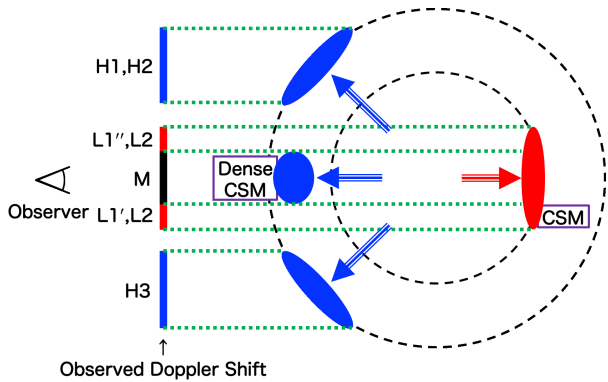}
 \end{center}
\caption{Schematic view of Kepler's SNR with region name, describing the observed ejecta properties. An observer is at the left. See text for more details.}\label{fig:punch}
\end{figure}

On the other hand, 
no shift in the Fe K$\alpha$ centroid in Region M is suggestive of that red- and blue-shifted ejecta are superimposed in this region. 
More specifically, based on the clumpy and diffuse appearances of blue- and red-shifted ejecta, 
blue-shifted ejecta corresponding to a structure seen in Region M in the Fe K$\alpha$ line flux image
may be superimposed on diffuse red-shifted ejecta widely distributing over Region L1, L2, and M. 
To confirm this idea, we fit the Fe K$\alpha$ line spectrum in Region E with two Gaussians. 
One is responsible for the red-shifted ejecta that has the same centroid energy and the line width as those of the Fe K$\alpha$ line in Region L2 (Gaussian 1) 
and the other is for additional blue-shifted ejecta (Gaussian 2). 
The spectrum is successfully reproduced with this model (figure \ref{fig:discussion_2_E} and best-fit values in table \ref{tab:discussion_2_E}). 
The derived $v_\mathrm{sight}$ for the additional blue-shifted component in Region M is comparable with those in Region H1--H3, 
or may be slower considering that Region M is located at the center of the remnant and less subject to projection effect. 
Figure \ref{fig:discussion_2_map} shows the O band map (0.5 -- 0.7 keV band) contours, 
which represent the CSM distribution (\cite{Katsuda15}), on the MPE map (figure \ref{fig:F_M_S}(b)). 
A clear morphological coincidence between the dense CSM structure and the shape of ``no-line-shift region'' (Region M) is evident, 
suggesting that the blue-shifted ejecta in Region M would have interacted with the dense CSM on the near side of the SNR and be decelerated. 
This scenario is supported by the fact that the optical filaments in this region are blue-shifted (\cite{Blair91,Blair07}). 
\citet{Blair07} also showed the CSM located in Region L1$^\prime$ are red-shifted 
and the velocity there is close to that at Region M. 
This fact suggests the red-shifted ejecta in Region L1$^\prime$ would have interacted with the CSM on the far side in contrast to Region M.
Figure \ref{fig:punch} shows a schematic drawing of the Fe ejecta properties in Kepler's SNR, which summarizes our understating.

One of the possible reasons that explains the Fe ejecta asymmetricity would be the SNR interaction with the asymmetric ambient medium. 
Recent three-dimensional numerical studies have succeeded in modeling a similar asymmetric structure to that of Kepler's SNR (e.g., \cite{Chiotellis12,Toledo-Roy14}). 
\citet{Toledo-Roy14} calculated the SNR models assuming a runaway progenitor scenario proposed by \citet{Bandiera87}. 
The model assumed that the SN explosion occurred inside an asymmetric CSM density distribution produced by the strong wind of the companion AGB star 
that was running through the ambient Galactic medium with a velocity of 280 km s$^{-1}$. 
As a result, the asymmetric bright X-ray structure around the central region as seen in Kepler's SNR was produced in the simulations (see Figure 10 in \cite{Toledo-Roy14}). 
However, the authors have argued such a structure is associated with the interaction of the supernova shockwave and the AGB wind. 
In our results, the Fe distributions in both spatial and velocity spaces have no strong correlation with the CSM distributions except for Region E (see also \cite{Burkey13}). 
Therefore, the CSM asymmetry seems not to be a predominant origin of the asymmetric distribution of the Fe ejecta.

\citet{Burkey13} argued that the ``shadow'' in Fe cast by the companion star might be able to produce the Fe ejecta asymmetricity. 
Recently, \citet{Garcia-Senz12} predicted a hidden hole that remains during centuries in type Ia SNRs caused by the interaction of the ejected materials with the companion star (shadow effect) 
using the three-dimensional hydrodynamical simulations (see also \cite{Gray16}). 
The model shows a cone-like hole structure during the young SNR stage, which would make SNR asymmetry in both spatial and velocity space. 
Also, it is notable that the model predicts the Rayleigh-Taylor instabilities at the outer edge of the hole develop faster than the average growth rate in the other unstable regions of the SNR models. 
Such an effect can push the ejecta that have high velocity and density close to the forward shock, and then the model might be able to explain the Fe ejecta asymmetricity in Kepler's SNR. 
For example, we showed the red-shifted Fe ejecta are located at the center and they are relatively uniformly distributed structures resulting in the diffuse appearance without distinct shape. 
On the other hand, the blue-shifted Fe ejecta are located at larger off-angle axis from the center and they have the clear shapes. 
The structure in which the blue-shift components surrounds the red-shifted component may indicate the projected hole structure if the hole faces us.

Another possible cause of the Fe ejecta asymmetricity would be an asymmetric explosion of the progenitor star. In the core of the SN Ia, a large amount of $^{56}$Ni ($\sim$0.6 $M_{\odot}$) are synthesized, 
and their asymmetric distribution during the SN explosion due to ignition process and/or some instabilities is expected (e.g., \cite{Maeda10,Seitenzahl13}). 
\citet{Yamaguchi17} have focused on the ``Fe knot'' located along the eastern rim of the Tycho's SNR, and attempted to explain the feature based on the explosion mechanism. 
This approach for understanding the mechanism of the Fe-knot formation would be also useful for our discussion of the Fe ejecta asymmetricity. 
For example, they discussed asymmetric Fe distribution using the N100 model of \citet{Seitenzahl13}. 
The model predicted some Fe clumps appear on the outer layer of SNe Ia during the SN explosion. 
The Fe clumps made in the initial stage are almost freely expanding and are thought to survive until the young SNR stage. 
Such an effect might make the Fe distribution as seen in Kepler's SNR.  
In order to conclude that such an asymmetric explosion is the origin of the Fe ejecta asymmetricity in the remnant, comparing the distribution with that of the other stable iron-peak elements (e.g., Cr, Mn, Ni) would be really needed. 
The Fe clumps ejected asymmetrically are thought to be synthesized in the neutron-rich Nuclear Statistical Equilibrium (n-NSE) layer and be scattered by some instabilities during the explosion. 
Therefore, the other n-NSE burning products (e.g, Mn and Ni) must be produced in the same region (see figure 13 in \cite{Yamaguchi17}). 
However, in present, the total X-ray counts in each region are too poor to estimate the element mass from the week lines (see figure \ref{fig:discussion_2_E}). 
In the future, even-longer-exposure observations by Chandra or XMM-Newton will be able to reveal it.

\section{Conclusions}\label{sec:conclusion}
The kinematic asymmetry of iron-peak elements in the type Ia SNe is the key to understanding the explosion mechanism,  
and then studying ejecta motion of SNRs is useful for approaching it. 
In order to know the expansion structure of Fe ejecta of Kepler's SNR, 
we analyzed the Fe K band images and spectra using the Chandra 741.0 ksec observation. 
From the radial profiles of the flux and centroid ratio of Fe K$\beta$/K$\alpha$, 
we resolved the mixture of multi $n_\mathrm{e} t$ components 
and found that the reverse shock is now propagating from the edge of the remnant to the center. 
Combining this estimation of the $n_\mathrm{e} t$ and the Fe K$\alpha$ centroid of each regions picked up from the mean photon energy map, 
we found that there are the blue-shifted ejecta with the velocity of $\sim$3,000 km s$^{-1}$ 
whereas other ejecta near the center are red-shifted with the velocity of $\sim$2,000 km s$^{-1}$. 
At this time, it is difficult to conclude what the physical process forming the asymmetric Fe motion is. 
Our results favor the ``shadow'' of the companion star 
and/or the asymmetric explosion of the progenitor star as the origin of the Fe ejecta asymmetricity in the remnant. 
The asymmetric distribution of the CSM also could make the Fe ejecta asymmetricity, 
however it hardly explains no strong correlation between the Fe and CSM distributions. 
In order to reveal the formation of the Fe ejecta asymmetricity in Kepler's SNR, 
we believe there is still much that the current X-ray observatories (e.g., Chandra, XMM-Newton) can do in the future. 
In particular, investigating more detailed kinematics on the Fe and the other iron-peak elements will access the issues.
We hope that the additional deep observations of type Ia SNRs like Kepler's SNR will be planed in coming cycles.

\begin{ack}
We thank the anonymous referee for comments to improve the manuscript. 
We also thank Kazuhiro Nakazawa, Satoru Katsuda, Sangwook Park, Asami Hayato, John P. Hughes, and Yoshihiro Furuta for helpful suggestion and discussion. 
T.K. is supported by the Advanced Leading Graduate Course for Photon Science (ALPS) in the University of Tokyo, 
and T.S is supported by the Special Postdoctoral Researchers Program in RIKEN. 
This work is partly supported by the Japan Society for the Promotion of Science (JSPS) KAKENHI Grant Number 15K051017 and 16H03983. 
\end{ack}

\appendix 
\section*{Radial profile in the northern and the southern halves}
\begin{figure*}[!ht]
 \begin{center}
  \includegraphics[width=16cm]{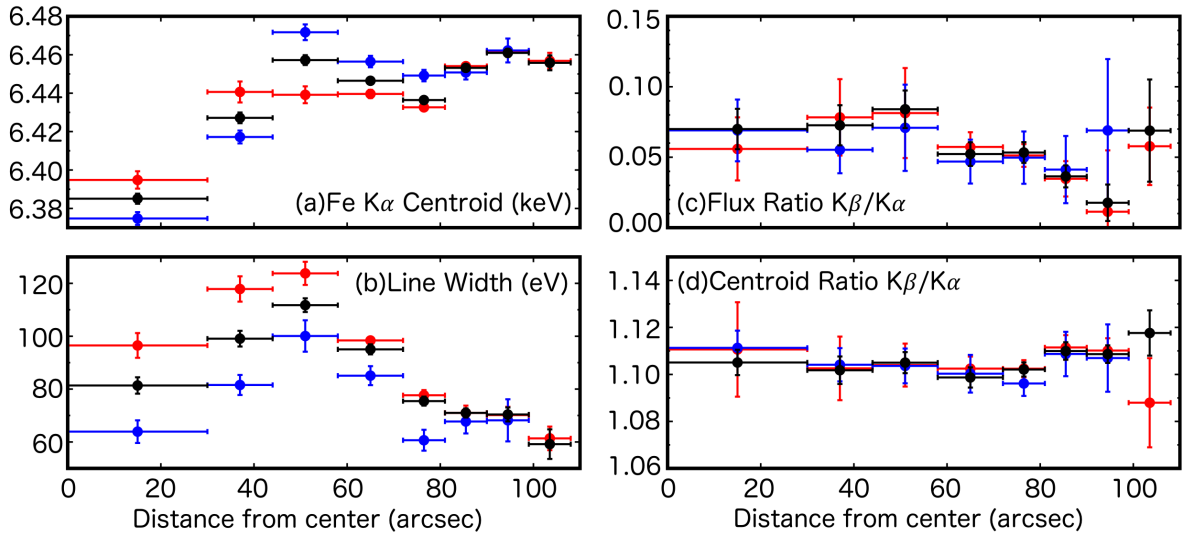}
 \end{center}
\caption{Result of the radial profiles of the north (red) and the south (blue). Black points are the same as figure \ref{fig:result_radial}. Error bars represent the 1$\sigma$ confidence level. The spectrum in the south of Region 8 cannot be analyzed due to the lack of statistics.}\label{fig:result_ns}
\end{figure*}

In addition to non-biased radial profiles in section \ref{sec:radial}, we also investigate radial profiles in northern/southern halves. 
We fit all spectra with the same method as section \ref{sec:radial} and the results are given in figure \ref{fig:result_ns}. 
Although the Fe K$\alpha$ centroid and line width are partly different, the flux ratio and the centroid ratio K$\beta$/K$\alpha$ are consistent in both halves. 
The difference would only be due to the Doppler motion in the line-of-sight direction, not the ionization effect. 
As a whole, the tendency is same as the result in section \ref{sec:radial}, 
that the center regions have smaller Fe K$\alpha$ centroid and smaller line width, supporting the asymmetric expansion. 

The northern/southern difference in the reverse shock dynamics might not be necessary for a runaway progenitor model (e.g., \cite{Bandiera87}). 
\citet{Chiotellis12} calculated the models for Kepler's SNR 
assuming that the progenitor system was a symbiotic binary moving toward the northwest with a velocity of 250 km s$^{-1}$. 
The models produce an asymmetric wind bubble around the progenitor. 
Then, the forward shock interaction with the wind bubble begins $\sim$300 yr after the explosion when the forward shock was at a radius of $\sim$2--3 pc. 
In this case, the northern/southern asymmetry of the forward shocks grows with time after the encounter. 
However, the growth rate of the asymmetry on the reverse shocks seems to be much lower than that of the forward shocks (see figure 5 in \cite{Chiotellis12}). 
This might support nonsignificant detection of the difference in the northern/southern radial dependences of the ionization states.


\begin{thebibliography}{}
\bibitem[Baade(1943)]{Baade43} Baade, W.\ 1943, \apj, 97, 119
\bibitem[Bamba et al.(2005)]{Bamba05} Bamba, A., Yamazaki, R., Yoshida, T., Terasawa, T., \& Koyama, K.\ 2005, \apj, 621, 793 
\bibitem[Bandiera(1987)]{Bandiera87} Bandiera, R.\ 1987, \apj, 319, 885
\bibitem[Bautz et al.(1998)]{Bautz98} Bautz, M.~W., et al.\ 1998, \procspie, 3444, 210
\bibitem[Blair et al.(1991)]{Blair91} Blair, W.~P., Long, K.~S., \& Vancura, O.\ 1991, \apj, 366, 484
\bibitem[Blair et al.(2007)]{Blair07} Blair, W.~P., Ghavamian, P., Long, K.~S., Williams, B.~J., Borkowski, K.~J., Reynolds, S.~P., \& Sankrit, R.\ 2007, \apj, 662, 998
\bibitem[Burkey et al.(2013)]{Burkey13} Burkey, M.~T., Reynolds, S.~P., Borkowski, K.~J., \& Blondin, J.~M.\ 2013, \apj, 764, 63
\bibitem[Cassam-Chena{\"i} et al.(2004)]{Cassam04} Cassam-Chena{\"i}, G., Decourchelle, A., Ballet, J., Hwang, U., Hughes, J.~P., \& Petre, R.\ 2004, \aap, 414, 545
\bibitem[Chiotellis et al.(2012)]{Chiotellis12} Chiotellis, A., Schure, K.~M., \& Vink, J.\ 2012, \aap, 537, A139
\bibitem[Furuzawa et al.(2009)]{Furuzawa09} Furuzawa, A., et al.\ 2009, \apjl, 693, L61
\bibitem[Garc{\'{\i}}a-Senz et al.(2012)]{Garcia-Senz12} Garc{\'{\i}}a-Senz, D., Badenes, C., \& Serichol, N.\ 2012, \apj, 745, 75
\bibitem[Gray et al.(2016)]{Gray16} Gray, W.~J., Raskin, C., \& Owen, J.~M.\ 2016, \apj, 833, 62
\bibitem[Hayato et al.(2010)]{Hayato10} Hayato, A., et al.\ 2010, \apj, 725, 894
\bibitem[Iwamoto et al.(1999)]{Iwamoto99} Iwamoto, K., Brachwitz, F., Nomoto, K., Kishimoto, N., Umeda, H., Hix, W.~R., \& Thielemann, F.-K.\ 1999, \apjs, 125, 439
\bibitem[Kasen et al.(2009)]{Kasen09} Kasen, D., R{\"o}pke, F.~K., \& Woosley, S.~E.\ 2009, \nat, 460, 869
\bibitem[Katsuda et al.(2008)]{Katsuda08} Katsuda, S., Tsunemi, H., Uchida, H., \& Kimura, M.\ 2008, \apj, 689, 225-230
\bibitem[Katsuda et al.(2015)]{Katsuda15} Katsuda, S., et al.\ 2015, \apj, 808, 49
\bibitem[Kerzendorf et al.(2014)]{Kerzendorf14} Kerzendorf, W.~E., Childress, M., Scharw{\"a}chter, J., Do, T., \& Schmidt, B.~P.\ 2014, \apj, 782, 27
\bibitem[Kuhlen et al.(2006)]{Kuhlen06} Kuhlen, M., Woosley, S.~E., \& Glatzmaier, G.~A.\ 2006, \apj, 640, 407
\bibitem[Livne et al.(2005)]{Livne05} Livne, E., Asida, S.~M., \& H{\"o}flich, P.\ 2005, \apj, 632, 443
\bibitem[Maeda et al.(2010)]{Maeda10} Maeda, K., et al.\ 2010, \nat, 466, 82
\bibitem[Maeda \& Terada(2016)]{Maeda16} Maeda, K., \& Terada, Y.\ 2016, International Journal of Modern Physics D, 25, 1630024
\bibitem[Maoz et al.(2014)]{Maoz14} Maoz, D., Mannucci, F., \& Nelemans, G.\ 2014, \araa, 52, 107
\bibitem[Park et al.(2013)]{Park13} Park, S., et al.\ 2013, \apjl, 767, L10
\bibitem[Reynolds et al.(2007)]{Reynolds07} Reynolds, S.~P., Borkowski, K.~J., Hwang, U., Hughes, J.~P., Badenes, C., Laming, J.~M., \& Blondin, J.~M.\ 2007, \apjl, 668, L135
\bibitem[Sankrit et al.(2005)]{Sankrit05} Sankrit, R., Blair, W.~P., Delaney, T., Rudnick, L., Harrus, I.~M., \& Ennis, J.~A.\ 2005, Advances in Space Research, 35, 1027
\bibitem[Sankrit et al.(2016)]{Sankrit16} Sankrit, R., Raymond, J.~C., Blair, W.~P., Long, K.~S., Williams, B.~J., Borkowski, K.~J., Pantnaude, D.~J., \& Reynolds, S.~P.\ 2016, \apj, 817, 36
\bibitem[Sato \& Hughes(2017a)]{Sato17a} Sato, T., \& Hughes, J.~P.\ 2017a, \apj, 840, 112
\bibitem[Sato \& Hughes(2017b)]{Sato17b} Sato, T., \& Hughes, J.~P.\ 2017b, \apj, 845, 167
\bibitem[Seitenzahl et al.(2013)]{Seitenzahl13} Seitenzahl, I.~R., et al.\ 2013, \mnras, 429, 1156
\bibitem[Toledo-Roy et al.(2014)]{Toledo-Roy14} Toledo-Roy, J.~C., Esquivel, A., Vel{\'a}zquez, P.~F., \& Reynoso, E.~M.\ 2014, \mnras, 442, 229
\bibitem[Vink(2008)]{Vink08} Vink, J.\ 2008, \apj, 689, 231-241
\bibitem[Vink(2017)]{Vink17} Vink, J.\ 2017, Handbook of Supernovae, ISBN 978-3-319-21845-8.~Springer International Publishing AG, 2017, p.~139, 139 
\bibitem[Whelan \& Iben(1973)]{Whelan73} Whelan, J., \& Iben, I., Jr.\ 1973, \apj, 186, 1007 
\bibitem[Williams et al.(2012)]{Williams12} Williams, B.~J., Borkowski, K.~J., Reynolds, S.~P., Ghavamian, P., Blair, W.~P., Long, K.~S., \& Sankrit, R.\ 2012, \apj, 755, 3
\bibitem[Williams et al.(2017)]{Williams17} Williams, B.~J., et al.\ 2017, \apj, 842, 28
\bibitem[Yamaguchi et al.(2014)]{Yamaguchi14} Yamaguchi, H., et al.\ 2014, \apj, 780, 136
\bibitem[Yamaguchi et al.(2015)]{Yamaguchi15} Yamaguchi, H., et al.\ 2015, \apjl, 801, L31
\bibitem[Yamaguchi et al.(2017)]{Yamaguchi17} Yamaguchi, H., Hughes, J.~P., Badenes, C., Bravo, E., Seitenzahl, I.~R., Mart{\'{\i}}nez-Rodr{\'{\i}}guez, H., Park, S., \& Petre, R.\ 2017, \apj, 834, 124
\end{thebibliography}
\end{document}